\documentclass[12pt,a4paper]{article}%
\usepackage{amsmath,amssymb,amsfonts}
\usepackage{graphicx,epsfig}
\usepackage{hyperref}
\usepackage{color}
\usepackage{amsmath}
\usepackage{amsfonts}
\usepackage{amssymb}
\usepackage{graphicx}
\usepackage{slashed}%
\numberwithin{equation}{section}
\begin{document}
\begin{titlepage}
\begin{flushright}
\end{flushright}
\vskip 1.0cm
\begin{center}
{\Large \bf Signatures of new physics at 14 TeV} \vskip 1.5cm
{\large Riccardo Barbieri}\\[1cm]
{\it Scuola Normale Superiore and INFN, Piazza dei Cavalieri 7, I-56126 Pisa, Italy} \\[5mm]
\vskip 2.0cm \abstract{I give an overview of the ideas and of the problems that orient the expectations for new physics at the Large Hadron Collider and, whenever I can, I describe the corresponding signals.}
\end{center}
\end{titlepage}

\section{A  {\it road map}}

The Large Hadron Collider will make the first 
 thorough exploration of the energy range at or well above  the Fermi scale,
$G_F^{-1/2}$, one of the two established fundamental scales in particle physics, the other being $\Lambda_{QCD}$. 
This is enough to make  the emergence of new phenomena highly plausible, whose description would require a 
revision of the Standard Model (SM) of elementary particles. Which new phenomena or, in the standard jargon, which {\it new physics}? The theoretical proposals are so many and so diverse that an impression of great confusion may easily be generated especially in a young person just approaching the field, as there are many present here.
The aim of this talk is to help correcting this impression. I try to  do this by describing a sort of {\it road map} that I will use myself to follow the flow of the experimental data with the focus on possible new phenomena. To the extent that this is possible, and to the best of my knowledge, in each case 
I will correspondingly indicate the relevant experimental signatures. 

In my view the main problems and the main ideas that orient the expectations for new physics at the LHC are the following (not in order of preference, see below):
\begin{itemize}
\item {\it Higgsless models}
\end{itemize}
That there be no Higgs boson is implausible. Yet, since no Higgs boson has been directly seen so far, one may want to take the conservative view that  no Higgs boson indeed exists. I would not give much weight to this possibility, were it not for the relatively recent suggestion\cite{SekharChivukula:2001hz,  Csaki:2003dt, Barbieri:2003pr} that part of the role of the Higgs boson could be played, in a potentially {\it calculable} way, by appropriate vectors, as we shall see.
\begin{itemize}
\item {\it The naturalness problem of the Fermi scale}
\end{itemize}
In a nutshell the famous naturalness problem\cite{Gildener:1976ai, Weinberg:1978ym} of the Fermi scale amounts to make the following hypothesis. There is a neat, not accidental reason that explains why short distance physics, whatever it may be, does not disturb the beautiful agreement of the Standard Model with the data. This is guaranteed if the short distance physics has in its infrared spectrum a naturally light Higgs boson. This highly motivated hypothesis remains the best theoretical reason for expecting new physics to show up at the LHC. We know of two possible explanations, although at different level of consistency, for a naturally light Higgs boson. One rests on {\it supersymmetry} \cite{Veltman:1976rt, Witten:1981nf, Dimopoulos:1981zb}. The other sees the Higgs doublet as a {\it pseudo-Goldstone} boson of a suitably broken global symmetry\cite{Kaplan:1983fs}. I generically refer in the following to this second possibility as the {\it composite} Higgs boson picture. Other names used in the literature in related  contexts are {\it Little Higgs} \cite{ArkaniHamed:2001nc} or {\it holographic} models\cite{Contino:2003ve, Agashe:2004rs}. This  interpretation  may be related with the existence of a compactified extra dimension, either in a true or in a metaphorical sense (via the so called AdS/CFT correspondence\cite{Maldacena:1997re}). 
\begin{itemize}
\item {\it Dark matter: a numerical coincidence}
\end{itemize}
To discover at the LHC an elementary constituent for Dark Matter (DM), seen in astrophysical and cosmological observations, would be a triumph for physics. The reasons we have to think that this might be the case rest on a numerical coincidence which  will be worth recalling\cite{Lee:1977ua, Goldberg:1983nd}. In turn this suggests the usefulness of taking a broad view when considering possible related signatures.
\begin{itemize}
\item {\it The $G_F^{-1/2}/M_{Pl}$ hierarchy as a manifestation of extra dimensions}
\end{itemize}
It is appealing to think that the huge difference between the Fermi and the Planck scales may be a manifestation of a suitably compactified extra dimension, one or more\cite{Antoniadis:1990ew,ArkaniHamed:1998rs,Randall:1999ee}. If this is true, significant gravitational phenomena could actually take place in particle physics experiments already at energies not too far from the Fermi scale itself and therefore potentially visible at the LHC. I find this possibility less likely than for any of the previous cases, which is why I shall not discuss it further. Nevertheless, this same scenario might be in the background of some of the more concrete possibilities mentioned above.

For reasons of time I shall not discuss the discovery potential at the LHC of new physics by measurements of flavour physics. Such potential exists, though, as exemplified by the case of the $B_s \rightarrow l^+ l^-$ decay.

\section{Higgsless models: a conservative view}
\label{higgsless}

In a new indefinite sector that replaces the Higgs doublet of the SM some dynamics breaks a global $SU(2)_L\times SU(2)_R \times U(1)_{B-L}$ symmetry down to $SU(2)_{L+R} \times U(1)_{B-L}$. At the same time the gauge group $SU(2)_L \times U(1)_Y$ gets broken down to $U(1)_{em}$ as desired. There are ways to describe this in a manifestly gauge invariant way  or even to make the dynamics explicit. The generic situation that results, however, either leaves many unknown parameters, like in the so-called Electroweak Chiral Lagrangian\cite{Appelquist,Longhitano}, or, when calculable, it is hardly compatible with the ElectroWeak Precision Tests (EWPT)\cite{Golden:1990ig,Peskin:1991sw, Altarelli:1991fk}, like in standard Technicolour\cite{Weinberg:1979bn, Susskind:1978ms}. It looks to me that more recent studies (models) have not substantially changed this situation. A feature that has emerged, though, which may deserve attention has to do with the unitarity problem of Higgsless models.

It is known since the seventies that in the SM with massive W and Z - call them collectively V - but without a Higgs boson, the $VV$ scattering amplitudes saturate unitarity already at a center of mass energy of $1\div 1.5$ TeV\cite{Lee:1977eg}. It is possible, however, to write down models\cite{Csaki:2003dt,Barbieri:2003pr}, either based on a compactified extra dimension or in some {\it deconstructed} version of them\cite{ArkaniHamed:2001ca,Hill:2000mu}, where the exchange in the $VV$ amplitude of heavy vectors - denoted by $\hat{V}$ - can prevent it from growing too fast\footnote{For an early model with a single heavy vector see (\cite{Casalbuoni:1985kq,Casalbuoni:1986vq,Bando:1987br}).}. In this way the saturation of unitarity of  $VV$ scattering can be postponed in a calculable way to  energies higher than $1\div 1.5$ TeV. The heavy $\hat{V}$s can be the beginning of a tower of states, all with the same quantum numbers, hence the possible name for them of Kaluza Klein (KK) vectors. Although  real calculability is achieved where the consistency with the EWPT gets  more problematic\cite{Barbieri:2003pr, Barbieri:2004qk},  this physical mechanism of keeping unitarity under partial control deserves attention. For this reason I summarize in Table \ref{heavyV} the main properties of the KK $\hat{V}$s, with the caveat that some of these properties are model dependent. In this Table $g_S$ is a strongish coupling, say $g_S \approx 3 \div 4$ or even bigger, whereas $g$ is the standard weak coupling. The scale $v$, which is not the vacuum expectation value of a Higgs field, is nevertheless still related to the W-mass in the usual way, $v = \sqrt{2} m_W/g=175$ GeV.

The KK vectors can be searched at the LHC via vector boson fusion, $q q \rightarrow q q~\hat{V}$, or by direct production, $q\bar{q} \rightarrow \hat{V}$, although through a suppressed coupling\cite{Birkedal:2004au,He:2007ge, Agashe:2007ki}. In turn $\hat{V}$ will decay into a pair of V bosons or into a pair or third generation quarks. A preliminary study of 
$qq \rightarrow qq~\hat{V}\rightarrow qqWZ \rightarrow qq~jj~ll$ can be found in Ref. \cite{Allanach:2006fy}. The decay of the KK vector into a pair of light fermions, like $\mu^+\mu^-$, is probably useless because of the small branching ratio. A high luminosity looks in any event mandatory for these searches (see below).

\begin{table}[ptb]%
\begin{equation}%
\begin{array}
[c]{c|c|c}%
 & \text{Higgsless} & \text{Composite}\\\hline
A(VV) & \approx s/v^2 & \approx s/f^2 \\[5pt]%
m_{\hat{V}} & \approx g_S v  & \approx g_S f \\
\hat{V} VV & g_S & g_S\\[5pt]%
f \bar{f} {\hat{V}} & \approx g (g/g_S) & \approx g (g/g_S)\\[5pt]%
Q_3 \bar{Q}_3 {\hat{V}}  & ? & \text{strongish}\\[5pt]
KK-\text{quarks} & - & \text{Yes, with}\approx \text{TeV mass}
\end{array}
\nonumber
\end{equation}
\caption{{\small Main phenomenological properties of Higgsless and Composite Higgs models.
$A(VV)$ is the $VV$-amplitude without the exchange of any KK vector, $\hat{V}$. $m_{\hat{V}}$ is the mass of the (first) KK vector. $\hat{V} VV$ is the strength of the triple coupling. $ f \bar{f} {\hat{V}}$ is the typical strength of the KK vector coupling to the light fermions. $Q_3 \bar{Q}_3{\hat{V}}$ is the coupling of the KK vector(s) to the third quark generation. $v = \sqrt{2} m_W/g=175$ GeV.}}
\label{heavyV}%
\end{table}

\section{{\it Composite} Higgs boson models}
\label{Composite}

For many reasons it is in any case far more likely that a Higgs boson exists. A way to protect its mass from large corrections is to make it an approximate Goldstone boson of a suitably broken global symmetry\cite{Kaplan:1983fs}. The set up is not very different from the one described at the beginning of  the previous Section, except for the fact that: i) The SM gauge group is fully inside the residual unbroken global group, $H$, and therefore remains also unbroken at a first stage; ii) (Some of) the Goldstone bosons associated with the breaking of the full global group $G$ down to $H$ must transform under the SM gauge group as the standard Higgs doublet and are called "Composite Higgs boson". This framework involves therefore two scales: the scale $f$ at which $G \rightarrow H$ and the usual vacuum expectation value $v$ of the Higgs field, with $f > v$. There exists a 5-dimensional variation of this scheme, where the Composite Higgs boson is interpreted as the fifth component of a vector in 5D. A simple example of a symmetry structure that works, with some advantageous phenomenological properties, is when $G = SO(5)\times U(1)_{B-L}$ and  $H = SO(4)\times U(1)_{B-L}$ \cite{Agashe:2004rs}.

The main features of this picture are the following (See Table \ref{heavyV}).
The scale $f$ cannot be too separated from $v$ unless one is ready to pay a fine tuning of order $(v/f)^2$. The $hVV$ couplings of the Higgs field $h$ with the W and the Z are suppressed, relative to the ones of the SM Higgs boson, by a factor 
$(1-v^2/f^2)^{1/2}$\cite{Giudice:2007fh, Barbieri:2007bh}. Therefore, as indicated in Table \ref{heavyV}, the amplitude $A(VV)$, even with the inclusion of the Higgs boson exchange, grows with $s$ as $\approx s/f^2$ and this growth must be partially compensated by KK vector exchanges. As usual the Higgs boson mass must be protected from large radiative corrections, especially the one due to the top exchange. In the Composite Higgs boson picture this happens because of the exchange of heavier vector-like quarks, which I shall call in the following KK quarks because they can occur in towers. Depending on the scheme, they can have charge 2/3 ($T$), 1/3 (B) or even exotic charges, like 5/3 ($X$). Finally, even Composite Higgs models, in their simplest versions and for not too large $f$ (i.e. a limited fine tuning), are not easily accommodated with the constraints of the EWPT\cite{Marandella:2005wd, Barbieri:2007bh, Carena:2007ua}. 
Perhaps the most plausible explanation of this problem is that we are ultimately dealing with a strong coupling theory and therefore we have a limited ability to do precise calculations.

The searches for KK vectors is similar here to the Higgsless case discussed above, with the difference that the KK vectors may be heavier (See Table \ref{heavyV}). If one takes  a  5-dimensional view, as I have indicated, there is also a pretty strong case for the existence of KK gluons decaying predominantly into $t\bar{t}$ pairs. The search for $pp \rightarrow \hat{g} \rightarrow t\bar{t}$ with sufficient luminosity looks promising\cite{Agashe:2006hk}. 

Relative to the KK vectors, the search for the KK quarks might actually be more fruitful, since  their masses, always pending the issue of the EWPT, might be significantly lower than the ones of the KK vectors. Denoting them collectively by $Q_{KK}$, their production by $q \bar{q} \rightarrow Q_{KK} \bar{Q}_{KK}$ and their subsequent decays, $Q_{KK} \rightarrow Q_3 V, Q_3 h$, may give rise to significant signals even with relatively moderate luminosities\cite{Dennis:2007tv, Contino:2008hi}.

\section{Supersymmetry}
\label{Susy}

Supersymmetry is the other possible explanation for a natural Fermi scale. Relative to the Composite Higgs picture it has at least the advantage of being straightforwardly compatible with the EWPT. There is a basic reason for this. Unlike the case of the Composite Higgs boson, the cancellation of the divergent contributions to the Higgs mass takes place between loops of particles that have different spin (top/stop, gauge-vectors/gauginos, etc.). Hence they cannot mix with each other and cannot produce tree level corrections to the EWPT. Needless to say it would be unreasonable not to mention 
the many other important and independent motivations that supersymmetry has in its own (gauge coupling unification\cite{Dimopoulos:1981yj, Ibanez:1981yh, Dimopoulos:1981zb} for one). Nevertheless it is supersymmetry as a solution of the naturalness problem of the Fermi scale that motivates its visibility at the LHC and this is what matters here.

The phenomenology of the Minimal Supersymmetric Standard Model (MSSM) with supersymmetry breaking parameters as  in minimal Supergravity, or mSUGRA\cite{Barbieri:1982eh, Chamseddine:1982jx, Hall:1983iz}, and a stable neutralino as the Lightest Supersymmetric Particle (LSP) is probably the most studied case of  physics beyond the SM. Gluino and squark pair production, with their subsequent chain decays ending into a neutralino LSP, gives rise to the characteristic missing energy signal generally accompanied by  jets and most often high $p_T$ leptons.
The conclusion that a very significant portion of the parameter space of mSUGRA can be successfully explored in this way at the LHC even with a relatively modest integrated luminosity is important and reassuring\cite{Ball:2007zza}. A limit of this analysis, on the other hand, is that it rests on the special  s-particle spectra  produced by the benchmark points of the  mSUGRA parameter space. A complementary analysis based on few s-particle masses and motivated by naturalness\footnote
{The stop is the s-particle with the strongest coupling to the Higgs boson (so that $m_Z^2 \approx m_{H_u}^2 \approx m_{\tilde{t}}^2$) and the gluino influences the stop mass via the strong gauge coupling (so that $m_{\tilde{t}}^2 \approx m_{\tilde{g}}^2$). I am allowing for  non-unified gaugino masses. $\tilde{t}_L, \tilde{b}_L, \tilde{t}_R$ and $\tilde{g}$ with masses in the 500 GeV range and all other s-particles  heavier, except for one neutralino, are consistent with flavour-physics constraints if the mixing angles in the coupling $\tilde{g} \bar{d}_L \tilde{d}_L$ are comparable to the ones of the CKM matrix.}
could take the gluino and, for simplicity, a single stop as the lightest s-particles apart from the neutralino LSP, $\chi_0$. In such a case the relevant searches would again be gluino and stop pair production with  dominant decays of the gluino into $t \tilde{t}$ or 
$q \bar{q} + \chi_0$ and of the stop into $t + \tilde{g}$ or $t + \chi_0$, depending on which is the lighter of the two, the stop or the gluino\footnote{The angular distributions in these decays depend on other parameters than the masses, but this, at least in a first stage of the analysis, is probably a negligible effect.}.

It is also true that mSUGRA need not be the end of the story. 
Letting  aside for the time being the issue of the Higgs boson system, there is at least the possibility that the supersymmetry breaking scale is low enough that the gravitino, rather than the neutralino, be the LSP. In this case $\chi_0$, if it is the next-to-lightest supersymmetric particle, would decay as $\chi_0 \rightarrow \gamma + g_{3/2}$ or, if allowed by phase space, as $\chi_0 \rightarrow Z + g_{3/2}$ or even $\chi_0 \rightarrow h + g_{3/2} \rightarrow b\bar{b} + g_{3/2}$ \cite{Fayet:1977vd, Fayet:1979yb, Dine:1994vc, Dimopoulos:1996vz}. In principle, for a special choice of the parameters, it is also  possible that these decays take place with a displaced vertex in the detector. I do not see this as a priority in the supersymmetry searches, though. More likely, in my mind, is the possibility that the gluino\cite{ArkaniHamed:2004fb,Giudice:2004tc} or the stop\cite{Barbieri:2000vh} have, for a reason or another, a long lifetime and behave in the detector as stable particles. It is again reassuring to see that $dE/dx$ and time-of-flight measurements can  recognize the corresponding "R-hadrons" in  an efficient way\cite{Rizzi:2007zz}.

\section{The supersymmetric Higgs bosons}
\label{susyHiggs}

The unsuccessful searches of the Higgs boson, standard or non standard, at LEP2 are a matter of concern for the supersymmetric picture, since we have always thought that supersymmetry requires at least one relatively light Higgs boson. There are two possible attitudes that one can take with respect to this problem, both defendable in my mind, which I  describe in turn.

The first attitude takes the MSSM as a strict guideline. As very well known, an upper bound holds in this case for the mass of one of the CP even scalars forming the supersymmetric Higgs boson system
\begin{equation}
m_h^2 \leq M_Z^2 \cos^2{2\beta} + \frac{3 m_t^4}{4 \pi^2 v^2} \log{\frac{m_{\tilde{t}}^2}{m_t^2}},
 \label{mhbound}%
\end{equation}
where $\tan{\beta}=v_2/v_1$ is the usual ratio of the two doublet vacuum expectation values, $m_{\tilde{t}}^2$ is an average stop mass squared and I have neglected for simplicity an effect proportional to the square of the so called $A_t$ term.
To comply with the LEP2 bound one considers a relatively large value of $\tan{\beta}$, so as to maximize the tree level contribution to (\ref{mhbound}), and a stop mass close to about 1 TeV, since the radiative term grows logarithmically with it. A large $\tan{\beta}$ could be suggested by interpreting the (theoretically uncertain) anomaly in the muon $g-2$ as a supersymmetric effect. The heavy stop, on the other hand, requires swallowing, e.g. in mSUGRA, a large contribution to the Z mass, typically 
$\Delta M_Z^2 \approx (2\div 3) m_{\tilde{t}}^2$, that has to be cancelled by an unpleasant fine tuning. If this is true, the lightest supersymmetric Higgs boson is just around the corner of the LEP2 bound, $m_h = 115 \div 120$ GeV, and it has very SM-like properties.

An alternative view, which I find motivated, gives weight to the following considerations. Even assuming, for good reasons indeed, that supersymmetry is relevant in nature, {\it no theorem} requires it to be visible at the LHC. For this to be the case, one needs a {\it maximally natural} Fermi scale with little fine tuning. Therefore, since the top, and so the stop, are the particles with the strongest coupling to the Higgs boson, it makes especially sense to insist on a moderate stop mass. Maybe (\ref{mhbound}) is not the key equation and  the supersymmetric Higgs boson system is not the one of the MSSM.
This view has been taken by many with different proposals. One that has received  attention is based on the Next-to-Minimal-Supersymmetric-Standard-Model (NMSSM) and wants a Higgs boson of mass around 100 GeV, in any case lower than the LEP2 bound of 115 GeV, which decays into two $\tau \bar{\tau}$ pairs\cite{Dermisek:2005gg, Chang:2005ht}. A reanalysis of the LEP data on this would in fact be welcome.

There are  different ways, however, of evading the LEP2 bound, also based on the NMSSM, which can be in some cases less fine tuned and may lead to very peculiar properties of the Higgs boson system. Here the crucial formula is the one that replaces (\ref{mhbound}) in the NMSSM
\begin{equation}
m^2_{hNMSSM} \leq M_Z^2 \cos^2{2\beta} + \lambda^2 v^2 \sin^2{2\beta} + \frac{3 m_t^4}{4 \pi^2 v^2} \log{\frac{m_{\tilde{t}}^2}{m_t^2}},
 \label{mhNMSSM}%
\end{equation}
with an extra tree-level contribution proportional to the square of the Yukawa coupling $\lambda S H_1 H_2$ between the two Higgs doublets and the singlet $S$. What counts therefore is the value of $\lambda$, with  two different cases that are interesting to consider.
\begin{itemize}
\item $\lambda (10~TeV) \leq 3$, 
\end{itemize}
so that  perturbative control of the EWPT is maintained. Remember that  the relevant expansion parameter is $\lambda^2/(4\pi)^2$ and that the presence of higher dimensional operators at a scale of 10 TeV or more does not disturb the perturbative calculation of the effects on the EWPT. By evolving $\lambda$ to the Fermi scale, which is where it counts for (\ref{mhNMSSM}), one has $\lambda(G_F^{-1/2}) \leq 2$. Since $\lambda$ gets non perturbative at relatively low energies, all this is a priori not  consistent with perturbative unification of the gauge coupling, unless one specifies in an appropriate way the change of regime that has to intervene  above about 10 TeV. 

 $\lambda(G_F^{-1/2}) \leq 2$ allows for a drastic departure from the usual supersymmetric Higgs picture\cite{Harnik:2003rs, Barbieri:2006bg}. The lightest Higgs boson can be as heavy as 300 GeV while being perfectly consistent with the EWPT and very similar to the SM Higgs boson. At the same time the heavier scalars can decay via the largish coupling $\lambda$ as, e.g., $h_2 \rightarrow h_1 h_1\rightarrow 4 V \rightarrow l^+ l^-~6 j$ or
$A_1 \rightarrow h_1 Z \rightarrow VV~Z \rightarrow l^+ l^-~4 j$. (As customary, $h$ is a scalar and $A$ a pseudoscalar.) The corresponding searches appear to be possible with a significant integrated luminosity\cite{Cavicchia:2007dp}. Since the lightest Higgs boson can be heavier than in the MSSM,  this picture has no fine tuning problem at all, even allowing for  relatively heavy stop and  gluino  (up to about 1 TeV and 2 TeV respectively) although probably still detectable at the LHC.

\begin{itemize}
\item $\lambda (M_{GUT}) \leq 3$, 
\end{itemize}
so that one keeps the consistency with manifest perturbative unification. Here the value of the evolved $\lambda$ at the low scale depends on the spectrum of matter between $M_{GUT}$ and $G_F^{-1/2}$. Since one does not want to disturb the success of supersymmetric gauge-coupling unification, this  matter at intermediate energies has to occur in full $SU(5)$ supermultiplets, like it happens in several motivated models. In this case $\lambda(G_F^{-1/2}) \approx 0.7\div 0.8$ can be attained\cite{Masip:1998jc, Barbieri:2007tu}.

$\lambda(G_F^{-1/2}) \approx 0.7\div 0.8$ also allows the scalar with the hZZ-coupling closest to the standard one to be above the LEP2 bound rather easily. In fact $m_h \approx 115\div 125$ GeV is possible {\it with a moderate stop mass}, say up to 300 GeV\cite{Barbieri:2007tu}.  It is also possible that this quasi-standard scalar is not the lightest member of the full Higgs boson system, which in the NMSSM is made of one charged state, 3 neutral scalars and two pseudoscalars. Among the pseudoscalars there can in fact be a quasi-Goldstone boson of an approximate Peccei Quinn symmetry, $A_1$\cite{Miller:2003ay,Schuster:2005py, Barbieri:2007tu}. In this quasi-symmetric limit,  $h$ can decay into a pair of stable neutralinos or as  $h \rightarrow A_1 A_1 \rightarrow b\bar{b}~b\bar{b}$, not the easiest mode to study at the LHC\cite{Carena:2007jk}.

\section{Dark Matter at the LHC}
\label{DM}

Suppose that there exists a neutral stable particle, $\chi$, of mass $m_\chi = O(G_F^{-1/2})$ that has been in equilibrium in the primordial hot plasma and, when the temperature of the plasma gets below its mass, its number density is reduced by  $\chi \chi \leftrightarrow f\bar{f}$, with $f$ a lighter standard particle. 
This is easily arranged by a discrete parity under which  $\chi \rightarrow -\chi$.
Up to corrections vanishing as $M_W/m_\chi$, the present relic energy density of the $\chi$ particle, in units of the critical density, is given by
\begin{equation}
\Omega_\chi h^2 = \frac{688 \pi^{5/2} T_\gamma ^3 x_f}{99 \sqrt{5 g_*} (H_0/h)^2 M_{Pl}^3 \sigma}
\approx 0.1 \frac{pb}{\sigma}
\label{omega}
\end{equation}
where $H_0$ is the present Hubble constant, $T_\gamma$ is the CMB temperature, $g_*$ is the number of effective degrees of freedom in the plasma at $T_f$ when the $\chi$ number density gets frozen, $x_f = m_\chi /T_f \approx 20\div 25$, and $\sigma$ is the thermal-averaged non-relativistic cross section for $\chi \chi \rightarrow f\bar{f}$. I have explicitly written down this formula to make evident that the final numerical result comes from a combination of many different physical constants. Now the remarkable and  famous coincidence\cite{Lee:1977ua, Goldberg:1983nd} is that, on one side, $\sigma \approx pb$ is the typical weak interaction cross section for a particle of mass $m_\chi = O(G_F^{-1/2})$ and, on the other side,  the observed energy density of cosmological Dark Matter is \cite{DMdata}
\begin{equation}
\Omega_{DM} h^2 = 0.113 \pm 0.009.
\label{omegaDM}
\end{equation}
This is enough to take seriously the possibility that a particle like $\chi$, generally called Weekly Interacting Massive Particle (WIMP), make the Dark Matter in the universe and can perhaps be discovered at the LHC.

As well know and evident from Section \ref{Susy}, a strongly  motivated candidate for this WIMP is the supersymmetric neutralino LSP, which has the potential advantage of being copiously produced in the chain decays of strongly interacting particles with a large production cross section.  Nevertheless, the above considerations suggest the usefulness of taking a broader point of view. I substantiate this statement by briefly describing two "minimal" examples of consistent Dark Matter candidates.

In the first one\cite{Deshpande:1977rw, Barbieri:2006dq} the SM is extended to include a second Higgs doublet $H_I$, $I$ for {\it inert}, with the following property: i)  it is not coupled to fermions because of an imposed $H_I \rightarrow - H_I$ parity to avoid any potential flavour problem; ii) it has a positive mass squared so that it gets no non-vanishing vacuum expectation value (hence the name of inert). This extra doublet leads to one charged and two neutral scalars, $H_0, A_0$, whose masses depend on their interactions with the standard Higgs doublet. The lightest of them is stable and, if neutral, makes a possible Dark Matter candidate. Its relic abundance has been studied and shown to be consistent with observations in a mass range between about 40 and 80 GeV and a small splitting, $\Delta m \approx 10$ GeV, with the heavier scalar, also neutral\cite{Barbieri:2006dq, Hambye:2007vf, Tytgat:2007cv}. The inert doublet can have a significant impact on the decay properties of the standard Higgs boson or, indirectly, on its mass range, as indicated by the EWPT\cite{Barbieri:2006dq}. At the LHC the detection of $pp \rightarrow A_0 H_0$, with the heavier scalar decaying into the lighter one plus a virtual Z, is very challanging\cite{Barbieri:2006dq,Cao:2007rm}.

 The second example\cite{Mahbubani:2005pt, D'Eramo:2007ga, Enberg:2007rp} makes use of fermions: a lepton-like vector doublet, $L = (\nu, E)$ and $L^c = (E^c, \nu^c)$ and a singlet $N$. Other than the covariant kinetic terms, the general Lagrangian that involves them is
 \begin{equation}
 \mathcal{L} = -\lambda L H N - \lambda^{\prime} L^c H^+ N + M_L L L^c + \frac{1}{2} M N^2 + h.c.
 \label{Lag}
 \end{equation}
 where $H$ is the standard Higgs doublet. After electroweak symmetry breaking the spectrum consists of one charged state, $E^\pm$, and three neutral Majorana fermions, $\nu_{1,2,3}$, the lightest of which can be a Dark Matter candidate. For some fixed values of the two Yukawa couplings $\lambda, \lambda^{\prime}$, even in this case the relic density has been studied and shown to be consistent with observations in a region of the $(M, M_L)$ plane. In such a region, one has also studied\cite{Enberg:2007rp} the expectations for a signal in  direct DM searches with bolometric detectors 
 and at the LHC for $pp \rightarrow E^\pm \nu_{2,3} \rightarrow W^\pm Z \nu_1 \nu_1 \rightarrow 3l + E_T$.
 In the LHC case, suitable cuts can lead to a discovery but the luminosity requirements are severe, to say the least. In general, I  believe that the lesson of these "minimal" models has to be kept in mind.

\section{Summary}

We expect that the LHC will unravel the physics of electroweak symmetry breaking by discovering the Higgs and/or new phenomena not included in the SM. 
This is based on the fact that the energy range at or well above the Fermi scale will be explored for the first time. To me this makes 
the situation at the LHC quite different from the one of the previous hadron colliders.
%, where the particles discovered and their main properties (the W and the Z at the $Sp\bar{p}S$ and the top at the TEVATRON) were largely anticipated. 
The LHC case is more open to surprises, suggesting that one should correspondingly take a broader point of view when talking of possible  signals of new physics. Nevertheless some possibilities stand up, which are, in my mind, the ones I have described. The related signals are summarized in Table \ref{summary}, where I also grossly indicate in each case, and to the best of my knowledge, the needed integrated luminosity for discovery. The tentative and biased character of this Table is evident.  It goes without saying that most of these entries are mutually exclusive.

\begin{table}[ptb]
\begin{equation}%
\begin{tabular}{||c|@{}c@{}||}
\hline
$\int L\, dt \leq 1 fb^{-1}$ &
\begin{tabular} {c}
\makebox [5cm] {mSUGRA}\\
\hline
\makebox [5cm] {$pp \rightarrow \tilde{g} \tilde{g}, \tilde{t} \tilde{t}$}\\
\hline
\makebox [5cm] {$\chi \rightarrow g_{3/2} + \gamma /Z/h$}\\
\hline
\makebox [5cm] {R-hadrons}\\
\end{tabular}\\
\hline
$\int L\, dt = 1 \div 30\,fb^{-1} $&
\begin{tabular} {c}
\makebox [5cm] {SM-like Higgs boson}\\
\hline
\makebox [5cm] {KK quarks}\\
\end{tabular}\\
\hline
$\int L\, dt > 30 fb^{-1}$ &
\begin{tabular} {c}
\makebox [5cm] {Susy Higgs boson system}\\
\hline
\makebox [5cm] {Minimal Dark Matter}\\
\hline
\makebox [5cm] {KK weak bosons}\\
\hline
\makebox [5cm] {KK gluons}\\
\end{tabular}\\
\hline
\end{tabular}
\nonumber
\end{equation}
\caption{{\small Summary of signals  as described in the text with a tentative estimate of the needed integrated luminosity for discovery. Most of these entries are mutually exclusive. The cases indicated under $\int L\, dt > 30 fb^{-1}$ may in fact be very challenging.}}
\label{summary}%
\end{table}

I want to conclude with a general remark. The physics of the Fermi scale is the physics of electroweak symmetry breaking, which  can be considered in many respects the current central question of particle physics and is  the focus of the activity at the LHC. At a somewhat deeper level and from a broader perspective I think that an equally, if not more, important question is the following: Which is the next relevant symmetry in particle physics, if any? 

The role of symmetries in describing the physics of the fundamental interactions does not have to be illustrated. Symmetries have been crucial in keeping the greatest economy in the number of principles and equations, which is the basic character of particle physics. Their enumeration, from Maxwell on, is unnecessary. The last one that has been experimentally established is the gauge symmetry of the SM. The assumption that symmetries will continue to play a central role in particle physics is implicit in all the considerations developed in  the previous pages. Such an assumption is being currently questioned in some circles. The LHC should shed light on this issue. 

\section*{{Acknowledgments}}

{This work is supported by the EU under RTN
contract MRTN-CT-2004-503369 and by the
MIUR under contract 2006022501. I thank 
Guido Altarelli, Alex Pomarol, Riccardo Rattazzi, Alessandro Strumia, Brando Bellazzini, Slava Rychkov, Alvise Varagnolo, Gian Giudice, Michelangelo Mangano, Guido Marandella, Michele Papucci, Lawrence Hall, Yasunori Nomura, Sergio Ferrara, Carlos Savoy, Duccio Pappadopulo, Gino, Isidori, Francesco D'Eramo, Leone Cavicchia, Roberto Franceschini
for many 
useful discussions and comments.}


\begin{thebibliography}{99}                                                                                               %

%\cite{SekharChivukula:2001hz}
\bibitem{SekharChivukula:2001hz}
  R.~Sekhar Chivukula, D.~A.~Dicus and H.~J.~He,
  %``Unitarity of compactified five dimensional Yang-Mills theory,''
  Phys.\ Lett.\  B {\bf 525} (2002) 175
  [arXiv:hep-ph/0111016].
  
  
%\cite{Csaki:2003dt}
\bibitem{Csaki:2003dt}
  C.~Csaki, C.~Grojean, H.~Murayama, L.~Pilo and J.~Terning,
  %``Gauge theories on an interval: Unitarity without a Higgs,''
  Phys.\ Rev.\  D {\bf 69}, 055006 (2004)
  [arXiv:hep-ph/0305237].
  
  %\cite{Barbieri:2003pr}
\bibitem{Barbieri:2003pr}
  R.~Barbieri, A.~Pomarol and R.~Rattazzi,
  %``Weakly coupled Higgsless theories and precision electroweak tests,''
  Phys.\ Lett.\  B {\bf 591}, 141 (2004)
  [arXiv:hep-ph/0310285].

%\cite{Gildener:1976ai}
\bibitem{Gildener:1976ai}
  E.~Gildener,
  %``Gauge Symmetry Hierarchies,''
  Phys.\ Rev.\  D {\bf 14} (1976) 1667.
  
  %\cite{Weinberg:1978ym}
\bibitem{Weinberg:1978ym}
  S.~Weinberg,
  %``Gauge Hierarchies,''
  Phys.\ Lett.\  B {\bf 82} (1979) 387.
  
%\cite{Veltman:1976rt}
\bibitem{Veltman:1976rt}
  M.~J.~G.~Veltman,
  %``Second Threshold In Weak Interactions,''
  Acta Phys.\ Polon.\  B {\bf 8} (1977) 475
  
  %\cite{Witten:1981nf}
\bibitem{Witten:1981nf}
  E.~Witten,
  %``Dynamical Breaking Of Supersymmetry,''
  Nucl.\ Phys.\  B {\bf 188} (1981) 513.
  
  %\cite{Dimopoulos:1981zb}
\bibitem{Dimopoulos:1981zb}
  S.~Dimopoulos and H.~Georgi,
  %``Softly Broken Supersymmetry And SU(5),''
  Nucl.\ Phys.\  B {\bf 193} (1981) 150.
  
  %\cite{Kaplan:1983fs}
\bibitem{Kaplan:1983fs}
  D.~B.~Kaplan and H.~Georgi,
  %``SU(2) X U(1) Breaking By Vacuum Misalignment,''
  Phys.\ Lett.\  B {\bf 136} (1984) 183.
  
  %\cite{ArkaniHamed:2001nc}
\bibitem{ArkaniHamed:2001nc}
  N.~Arkani-Hamed, A.~G.~Cohen and H.~Georgi,
  %``Electroweak symmetry breaking from dimensional deconstruction,''
  Phys.\ Lett.\  B {\bf 513} (2001) 232
  [arXiv:hep-ph/0105239].
  
 %\cite{Contino:2003ve}
\bibitem{Contino:2003ve}
  R.~Contino, Y.~Nomura and A.~Pomarol,
  %``Higgs as a holographic pseudo-Goldstone boson,''
  Nucl.\ Phys.\  B {\bf 671} (2003) 148
  [arXiv:hep-ph/0306259].
  
   %\cite{Agashe:2004rs}
\bibitem{Agashe:2004rs}
  K.~Agashe, R.~Contino and A.~Pomarol,
  %``The minimal composite Higgs model,''
  Nucl.\ Phys.\  B {\bf 719} (2005) 165
  [arXiv:hep-ph/0412089].
  
  %\cite{Maldacena:1997re}
\bibitem{Maldacena:1997re}
  J.~M.~Maldacena,
  %``The large N limit of superconformal field theories and supergravity,''
  Adv.\ Theor.\ Math.\ Phys.\  {\bf 2} (1998) 231
  [Int.\ J.\ Theor.\ Phys.\  {\bf 38} (1999) 1113]
  [arXiv:hep-th/9711200].
  
%\cite{Lee:1977ua}
\bibitem{Lee:1977ua}
  B.~W.~Lee and S.~Weinberg,
  %``Cosmological lower bound on heavy-neutrino masses,''
  Phys.\ Rev.\ Lett.\  {\bf 39} (1977) 165.
  
  %\cite{Goldberg:1983nd}
\bibitem{Goldberg:1983nd}
  H.~Goldberg,
  %``Constraint on the photino mass from cosmology,''
  Phys.\ Rev.\ Lett.\  {\bf 50} (1983) 1419.
  
  %\cite{Antoniadis:1990ew}
\bibitem{Antoniadis:1990ew}
  I.~Antoniadis,
  %``A Possible new dimension at a few TeV,''
  Phys.\ Lett.\  B {\bf 246} (1990) 377.
  
  %\cite{ArkaniHamed:1998rs}
\bibitem{ArkaniHamed:1998rs}
  N.~Arkani-Hamed, S.~Dimopoulos and G.~R.~Dvali,
  %``The hierarchy problem and new dimensions at a millimeter,''
  Phys.\ Lett.\  B {\bf 429} (1998) 263
  [arXiv:hep-ph/9803315].
  
  %\cite{Randall:1999ee}
\bibitem{Randall:1999ee}
  L.~Randall and R.~Sundrum,
  %``A large mass hierarchy from a small extra dimension,''
  Phys.\ Rev.\ Lett.\  {\bf 83} (1999) 3370
  [arXiv:hep-ph/9905221].
  
    %\cite{Appelquist}
\bibitem{Appelquist}
T. Appelquist and C. W. Bernard, Phys. Rev. D22 (1980) 200.

    %\cite{Longhitano}
\bibitem{Longhitano}
  A. C. Longhitano, Nucl. Phys. B188 (1981) 118. 
  
  %\cite{Golden:1990ig}
\bibitem{Golden:1990ig}
  M.~Golden and L.~Randall,
  %``RADIATIVE CORRECTIONS TO ELECTROWEAK PARAMETERS IN TECHNICOLOR THEORIES,''
  Nucl.\ Phys.\  B {\bf 361} (1991) 3.
  
  %\cite{Peskin:1991sw}
\bibitem{Peskin:1991sw}
  M.~E.~Peskin and T.~Takeuchi,
  %``Estimation of oblique electroweak corrections,''
  Phys.\ Rev.\  D {\bf 46} (1992) 381.
 
 %\cite{Altarelli:1991fk}
\bibitem{Altarelli:1991fk}
  G.~Altarelli, R.~Barbieri and S.~Jadach,
  %``Toward a model independent analysis of electroweak data,''
  Nucl.\ Phys.\  B {\bf 369} (1992) 3
  [Erratum-ibid.\  B {\bf 376} (1992) 444]
  
%\cite{Weinberg:1979bn}
\bibitem{Weinberg:1979bn}
  S.~Weinberg,
  %``Implications Of Dynamical Symmetry Breaking: An Addendum,''
  Phys.\ Rev.\  D {\bf 19} (1979) 1277.

  %\cite{Susskind:1978ms}
\bibitem{Susskind:1978ms}
  L.~Susskind,
  %``Dynamics Of Spontaneous Symmetry Breaking In The Weinberg-Salam Theory,''
  Phys.\ Rev.\  D {\bf 20} (1979) 2619.

  %\cite{Lee:1977eg}
\bibitem{Lee:1977eg}
  B.~W.~Lee, C.~Quigg and H.~B.~Thacker,
  %``Weak Interactions At Very High-Energies: The Role Of The Higgs Boson
  %Mass,''
  Phys.\ Rev.\  D {\bf 16} (1977) 1519.

%\cite{ArkaniHamed:2001ca}
\bibitem{ArkaniHamed:2001ca}
  N.~Arkani-Hamed, A.~G.~Cohen and H.~Georgi,
  %``(De)constructing dimensions,''
  Phys.\ Rev.\ Lett.\  {\bf 86} (2001) 4757
  [arXiv:hep-th/0104005].
  
  %\cite{Hill:2000mu}
\bibitem{Hill:2000mu}
  C.~T.~Hill, S.~Pokorski and J.~Wang,
  %``Gauge invariant effective Lagrangian for Kaluza-Klein modes,''
  Phys.\ Rev.\  D {\bf 64} (2001) 105005
  [arXiv:hep-th/0104035].
  
  %\cite{Casalbuoni:1985kq}
\bibitem{Casalbuoni:1985kq}
  R.~Casalbuoni, S.~De Curtis, D.~Dominici and R.~Gatto,
  %``Effective Weak Interaction Theory With Possible New Vector Resonance From A
  %Strong Higgs Sector,''
  Phys.\ Lett.\  B {\bf 155} (1985) 95.
  
  %\cite{Casalbuoni:1986vq}
\bibitem{Casalbuoni:1986vq}
  R.~Casalbuoni, S.~De Curtis, D.~Dominici and R.~Gatto,
  %``Physical Implications Of Possible J=1 Bound States From Strong Higgs,''
  Nucl.\ Phys.\  B {\bf 282} (1987) 235.

  %\cite{Bando:1987br}
\bibitem{Bando:1987br}
  M.~Bando, T.~Kugo and K.~Yamawaki,
  %``Nonlinear Realization and Hidden Local Symmetries,''
  Phys.\ Rept.\  {\bf 164} (1988) 217.
  
%\cite{Barbieri:2004qk}
\bibitem{Barbieri:2004qk}
  R.~Barbieri, A.~Pomarol, R.~Rattazzi and A.~Strumia,
  %``Electroweak symmetry breaking after LEP1 and LEP2,''
  Nucl.\ Phys.\  B {\bf 703} (2004) 127
  [arXiv:hep-ph/0405040].
  
  %\cite{Birkedal:2004au}
\bibitem{Birkedal:2004au}
  A.~Birkedal, K.~Matchev and M.~Perelstein,
  %``Collider phenomenology of the Higgsless models,''
  Phys.\ Rev.\ Lett.\  {\bf 94} (2005) 191803
  [arXiv:hep-ph/0412278].
  
  %\cite{He:2007ge}
\bibitem{He:2007ge}
  H.~J.~He {\it et al.},
  %``LHC Signatures of New Gauge Bosons in Minimal Higgsless Model,''
  arXiv:0708.2588 [hep-ph].
  
  %\cite{Agashe:2007ki}
\bibitem{Agashe:2007ki}
  K.~Agashe {\it et al.},
  %``LHC Signals for Warped Electroweak Neutral Gauge Bosons,''
  Phys.\ Rev.\  D {\bf 76} (2007) 115015
  [arXiv:0709.0007 [hep-ph]].
  
  %\cite{Allanach:2006fy}
\bibitem{Allanach:2006fy}
  See G. Azuelos, P-A. Delsart and J. Idarraga in
  B.~C.~Allanach {\it et al.},
  %``Les Houches 'Physics at TeV colliders 2005' Beyond the standard model
  %working group: Summary report,''
  arXiv:hep-ph/0602198.
  
  %\cite{Giudice:2007fh}
\bibitem{Giudice:2007fh}
  G.~F.~Giudice, C.~Grojean, A.~Pomarol and R.~Rattazzi,
  %``The Strongly-Interacting Light Higgs,''
  JHEP {\bf 0706}, 045 (2007)
  [arXiv:hep-ph/0703164].
  
  %\cite{Barbieri:2007bh}
\bibitem{Barbieri:2007bh}
  R.~Barbieri, B.~Bellazzini, V.~S.~Rychkov and A.~Varagnolo,
  %``The Higgs boson from an extended symmetry,''
  Phys.\ Rev.\  D {\bf 76}, 115008 (2007)
  [arXiv:0706.0432 [hep-ph]].
  
  %\cite{Marandella:2005wd}
\bibitem{Marandella:2005wd}
  G.~Marandella, C.~Schappacher and A.~Strumia,
  %``Little-Higgs corrections to precision data after LEP2,''
  Phys.\ Rev.\  D {\bf 72} (2005) 035014
  [arXiv:hep-ph/0502096].
  
  %\cite{Carena:2007ua}
\bibitem{Carena:2007ua}
  M.~S.~Carena, E.~Ponton, J.~Santiago and C.~E.~M.~Wagner,
  %``Electroweak constraints on warped models with custodial symmetry,''
  Phys.\ Rev.\  D {\bf 76} (2007) 035006
  [arXiv:hep-ph/0701055].
  
  %\cite{Agashe:2006hk}
\bibitem{Agashe:2006hk}
  K.~Agashe, A.~Belyaev, T.~Krupovnickas, G.~Perez and J.~Virzi,
  %``LHC signals from warped extra dimensions,''
  Phys.\ Rev.\  D {\bf 77} (2008) 015003
  [arXiv:hep-ph/0612015].
  
  %\cite{Dennis:2007tv}
\bibitem{Dennis:2007tv}
  C.~Dennis, M.~Karagoz Unel, G.~Servant and J.~Tseng,
  %``Multi-W events at LHC from a warped extra dimension with custodial
  %symmetry,''
  arXiv:hep-ph/0701158.
  
  %\cite{Contino:2008hi}
\bibitem{Contino:2008hi}
  R.~Contino and G.~Servant,
  %``Discovering the top partners at the LHC using same-sign dilepton final
  %states,''
  arXiv:0801.1679 [hep-ph].


%\cite{Dimopoulos:1981yj}
\bibitem{Dimopoulos:1981yj}
  S.~Dimopoulos, S.~Raby and F.~Wilczek,
  %``Supersymmetry And The Scale Of Unification,''
  Phys.\ Rev.\  D {\bf 24} (1981) 1681.
  
  %\cite{Ibanez:1981yh}
\bibitem{Ibanez:1981yh}
  L.~E.~Ibanez and G.~G.~Ross,
  %``Low-Energy Predictions In Supersymmetric Grand Unified Theories,''
  Phys.\ Lett.\  B {\bf 105} (1981) 439.
  
  %\cite{Barbieri:1982eh}
\bibitem{Barbieri:1982eh}
  R.~Barbieri, S.~Ferrara and C.~A.~Savoy,
  %``Gauge Models With Spontaneously Broken Local Supersymmetry,''
  Phys.\ Lett.\  B {\bf 119} (1982) 343.
  
  %\cite{Chamseddine:1982jx}
\bibitem{Chamseddine:1982jx}
  A.~H.~Chamseddine, R.~Arnowitt and P.~Nath,
  %``Locally Supersymmetric Grand Unification,''
  Phys.\ Rev.\ Lett.\  {\bf 49} (1982) 970.
  
%\cite{Hall:1983iz}
\bibitem{Hall:1983iz}
  L.~J.~Hall, J.~D.~Lykken and S.~Weinberg,
  %``Supergravity As The Messenger Of Supersymmetry Breaking,''
  Phys.\ Rev.\  D {\bf 27} (1983) 2359.

%\cite{Ball:2007zza}
\bibitem{Ball:2007zza}
  See, e.g., G.~L.~Bayatian {\it et al.}  [CMS Collaboration],
  %``CMS technical design report, volume II: Physics performance,''
  J.\ Phys.\ G {\bf 34} (2007) 995.
  
  
%\cite{Fayet:1977vd}
\bibitem{Fayet:1977vd}
  P.~Fayet,
  %``Mixing Between Gravitational And Weak Interactions Through The Massive
  %Gravitino,''
  Phys.\ Lett.\  B {\bf 70} (1977) 461.
  
%\cite{Fayet:1979yb}
\bibitem{Fayet:1979yb}
  P.~Fayet,
  %``Scattering Cross-Sections Of The Photino And The Goldstino (Gravitino) On
  %Matter,''
  Phys.\ Lett.\  B {\bf 86} (1979) 272.
  
 %\cite{Dine:1994vc}
\bibitem{Dine:1994vc}
  M.~Dine, A.~E.~Nelson and Y.~Shirman,
  %``Low-Energy Dynamical Supersymmetry Breaking Simplified,''
  Phys.\ Rev.\  D {\bf 51} (1995) 1362
  [arXiv:hep-ph/9408384].
  
  %\cite{Dimopoulos:1996vz}
\bibitem{Dimopoulos:1996vz}
  S.~Dimopoulos, M.~Dine, S.~Raby and S.~D.~Thomas,
  %``Experimental Signatures of Low Energy Gauge Mediated Supersymmetry
  %Breaking,''
  Phys.\ Rev.\ Lett.\  {\bf 76} (1996) 3494
  [arXiv:hep-ph/9601367].
  
  
  %\cite{ArkaniHamed:2004fb}
\bibitem{ArkaniHamed:2004fb}
  N.~Arkani-Hamed and S.~Dimopoulos,
  %``Supersymmetric unification without low energy supersymmetry and  signatures
  %for fine-tuning at the LHC,''
  JHEP {\bf 0506} (2005) 073
  [arXiv:hep-th/0405159].
  
  %\cite{Giudice:2004tc}
\bibitem{Giudice:2004tc}
  G.~F.~Giudice and A.~Romanino,
  %``Split supersymmetry,''
  Nucl.\ Phys.\  B {\bf 699} (2004) 65
  [Erratum-ibid.\  B {\bf 706} (2005) 65]
  [arXiv:hep-ph/0406088].
  
    %\cite{Barbieri:2000vh}
\bibitem{Barbieri:2000vh}
  R.~Barbieri, L.~J.~Hall and Y.~Nomura,
  %``A constrained standard model from a compact extra dimension,''
  Phys.\ Rev.\  D {\bf 63} (2001) 105007
  [arXiv:hep-ph/0011311].

  
  %\cite{Rizzi:2007zz}
\bibitem{Rizzi:2007zz}
  A.~Rizzi,
  ``Observability of R-hadrons at the LHC''
  
  %\cite{Dermisek:2005gg}
\bibitem{Dermisek:2005gg}
  R.~Dermisek and J.~F.~Gunion,
  %``Consistency of LEP event excesses with an h --> a a decay scenario and
  %low-fine-tuning NMSSM models,''
  Phys.\ Rev.\  D {\bf 73} (2006) 111701
  [arXiv:hep-ph/0510322]
  
  %\cite{Chang:2005ht}
\bibitem{Chang:2005ht}
  S.~Chang, P.~J.~Fox and N.~Weiner,
  %``Naturalness and Higgs decays in the MSSM with a singlet,''
  JHEP {\bf 0608} (2006) 068
  [arXiv:hep-ph/0511250].
  
  %\cite{Harnik:2003rs}
\bibitem{Harnik:2003rs}
  R.~Harnik, G.~D.~Kribs, D.~T.~Larson and H.~Murayama,
  %``The minimal supersymmetric fat Higgs model,''
  Phys.\ Rev.\  D {\bf 70} (2004) 015002
  [arXiv:hep-ph/0311349].
  
  %\cite{Barbieri:2006bg}
\bibitem{Barbieri:2006bg}
  R.~Barbieri, L.~J.~Hall, Y.~Nomura and V.~S.~Rychkov,
  %``Supersymmetry without a light Higgs boson,''
  Phys.\ Rev.\  D {\bf 75} (2007) 035007
  [arXiv:hep-ph/0607332].
  
  %\cite{Cavicchia:2007dp}
\bibitem{Cavicchia:2007dp}
  L.~Cavicchia, R.~Franceschini and V.~S.~Rychkov,
  %``Supersymmetry without a light Higgs boson at the LHC,''
  arXiv:0710.5750 [hep-ph].
  
  %\cite{Masip:1998jc}
\bibitem{Masip:1998jc}
  M.~Masip, R.~Munoz-Tapia and A.~Pomarol,
  %``Limits on the mass of the lightest Higgs in supersymmetric models,''
  Phys.\ Rev.\  D {\bf 57} (1998) 5340
  [arXiv:hep-ph/9801437].
  
%\cite{Barbieri:2007tu}
\bibitem{Barbieri:2007tu}
  R.~Barbieri, L.~J.~Hall, A.~Y.~Papaioannou, D.~Pappadopulo and V.~S.~Rychkov,
  %``An alternative NMSSM phenomenology with manifest perturbative
  %unification,''
  arXiv:0712.2903 [hep-ph].
  
  %\cite{Miller:2003ay}
\bibitem{Miller:2003ay}
  D.~J.~Miller, R.~Nevzorov and P.~M.~Zerwas,
  %``The Higgs sector of the next-to-minimal supersymmetric standard model,''
  Nucl.\ Phys.\  B {\bf 681} (2004) 3
  [arXiv:hep-ph/0304049].
  
  %\cite{Schuster:2005py}
\bibitem{Schuster:2005py}
  P.~C.~Schuster and N.~Toro,
  %``Persistent fine-tuning in supersymmetry and the NMSSM,''
  arXiv:hep-ph/0512189.
  
  
  %\cite{Carena:2007jk}
\bibitem{Carena:2007jk}
  M.~Carena, T.~Han, G.~Y.~Huang and C.~E.~M.~Wagner,
  %``Higgs Signal for h to aa at Hadron Colliders,''
  arXiv:0712.2466 [hep-ph].

  %\cite{DMdata}
\bibitem{DMdata}
G. Hinshaw et al., Astrophys. J. Suppl. 148 (2003) 135 [arXiv:astro-ph/0302217]; 
C. J. MacTavish et al., arXiv:astro-ph/0507503; A. G. Sanchez et al., 
arXiv:astro-ph/0507583. 

%\cite{Deshpande:1977rw}
\bibitem{Deshpande:1977rw}
  N.~G.~Deshpande and E.~Ma,
  %``Pattern Of Symmetry Breaking With Two Higgs Doublets,''
  Phys.\ Rev.\  D {\bf 18} (1978) 2574.
  
  
  %\cite{Barbieri:2006dq}
\bibitem{Barbieri:2006dq}
  R.~Barbieri, L.~J.~Hall and V.~S.~Rychkov,
  %``Improved naturalness with a heavy Higgs: An alternative road to LHC
  %physics,''
  Phys.\ Rev.\  D {\bf 74} (2006) 015007
  [arXiv:hep-ph/0603188].
  
  %\cite{Hambye:2007vf}
\bibitem{Hambye:2007vf}
  T.~Hambye and M.~H.~G.~Tytgat,
  %``Electroweak Symmetry Breaking induced by Dark Matter,''
  Phys.\ Lett.\  B {\bf 659} (2008) 651
  [arXiv:0707.0633 [hep-ph]].
  
  %\cite{Tytgat:2007cv}
\bibitem{Tytgat:2007cv}
  M.~H.~G.~Tytgat,
  %``The Inert Doublet Model : a new archetype of WIMP dark matter?,''
  arXiv:0712.4206 [hep-ph].
  
  %\cite{Cao:2007rm}
\bibitem{Cao:2007rm}
  Q.~H.~Cao, E.~Ma and G.~Rajasekaran,
  %``Observing the Dark Scalar Doublet and its Impact on the Standard-Model
  %Higgs Boson at Colliders,''
  Phys.\ Rev.\  D {\bf 76} (2007) 095011
  [arXiv:0708.2939 [hep-ph]].
  
  
%\cite{Mahbubani:2005pt}
\bibitem{Mahbubani:2005pt}
  R.~Mahbubani and L.~Senatore,
  %``The minimal model for dark matter and unification,''
  Phys.\ Rev.\  D {\bf 73} (2006) 043510
  [arXiv:hep-ph/0510064].
  
  %\cite{D'Eramo:2007ga}
\bibitem{D'Eramo:2007ga}
  F.~D'Eramo,
  %``Dark matter and Higgs boson physics,''
  Phys.\ Rev.\  D {\bf 76} (2007) 083522
  [arXiv:0705.4493 [hep-ph]].
  
  %\cite{Enberg:2007rp}
\bibitem{Enberg:2007rp}
  R.~Enberg, P.~J.~Fox, L.~J.~Hall, A.~Y.~Papaioannou and M.~Papucci,
  %``LHC and Dark Matter Signals of Improved Naturalness,''
  JHEP {\bf 0711} (2007) 014
  [arXiv:0706.0918 [hep-ph]].
  
  
  
  
\end{thebibliography}
\end{document}